\documentstyle[12pt]{article}
\def\beq{\begin{equation}}
\def\eeq{\end{equation}}
\def\bea{\begin{eqnarray}}
\def\eea{\end{eqnarray}}
\def\bmath{\begin{displaymath}}
\def\emath{\end{displaymath}}
\def\bq{\begin{quote}}
\def\eq{\end{quote}}
\def\slash#1{\setbox0=\hbox{$#1$}#1\hskip-\wd0\hbox to\wd0{\hss\sl/\/\hss}}
\begin{document}

\begin{flushright}
MZ-TH/93-04 \\
FTUV/93-24\\
May 1993
\end{flushright}

\begin{center}
{\bf{\Large UNIVERSALITY-BREAKING EFFECTS}} \\[0.45cm]
{\bf{\Large IN LEPTONIC {\boldmath $Z$} DECAYS}}\\[1.5cm]
{\large J.~Bernab\'eu}$^{\displaystyle \, \, (a) }$,
{\large J.G.~K\"orner}$^{\displaystyle \, \, (b)}$,
{\large A.~Pilaftsis}$^{\displaystyle \, \, (b) ,}$
\footnote[1]{Address after 1, Oct.~1993, Rutherford Appleton Laboratory,
Chilton, Didcot, Oxon,\\
{\em ENGLAND}. E-mail address: pilaftsis@vipmza.physik.uni-mainz.de},
{\large K.~Schilcher}$^{\displaystyle \, \, (b)}$
\\[0.7cm]
$^{(a)}$ Departament de Fisica Te\`orica, Univ.~de Valencia,\\
and IFIC, Univ.~de Val\`encia--CSIC,\\
E-46100 Burjassot (Val\`encia), {\em SPAIN}
\\[0.5cm]
$^{(b)}$ Institut f\"ur Physik, Johannes-Gutenberg Universit\"at,
Staudinger Weg 7,\\
W-6500 Mainz, {\em FRG}
\end{center}

\bigskip
\bigskip
\centerline {\bf ABSTRACT}

We analyze the possibility of universality violation in diagonal
leptonic decays of the $Z$ boson, in the context of interfamily
"see-saw" models. In a minimal extension of the Standard Model with
right-handed neutrino fields, we find that universality-breaking
effects increase quadratically with the heavy Majorana neutrino mass and
may be observed in the running $LEP$ experiments.\\[0.8cm]
PACS {\em numbers}: 12.15.J, 13.38, 14.80.E

\newpage

Heavy Majorana neutrinos with masses of few TeV have entered the domain of
cosmology and astrophysics, as possible candidates to account for the net
baryon number of the universe through lepton~($L$)-number violating
processes~\cite{cosmos}. On the other hand, if such heavy neutral leptons
are realized in nature, their existence may be discovered through
their production and the $L$-violating decay at present or future collider
machines~\cite{Lviol}. Another place of looking for new physics originating
from heavy neutrinos is the rare leptonic decays of $H^0$~\cite{AP} and
$Z$ particle~\cite{JB}. Since such decays are forbidden in the minimal
Standard Model ($SM$), they
constitute an interesting framework to constrain new physics beyond the $SM$.
Such rare decays~\cite{AP,KPS} have recently been analyzed in
a "see-saw"-type model~\cite{YGRS} with intergenerational mixings~\cite{ZPC}.
An interesting aspect of this minimal scenario is that the
Appelquist-Carazzone theorem~\cite{Appel} is not operative and
vertex-correction
diagrams with intermediate heavy Majorana neutrinos~($N_i$) show a
quadratic mass dependence, i.e.~$m^2_{N_i}/M^2_W$. This mass dependence is a
common feature for all theories based on the spontaneous-symmetry breaking
mechanism. For example, in the Feynman--'t Hooft gauge this mass dependence
can be seen to arise from the coupling of the would-be charged Goldstone
bosons to heavy fermions. Similar non-decoupling effects in the quark sector
have been extensively studied in the past for the flavor changing decays
$Z\to b\bar{s}$~\cite{BSbar} and the diagonal $Z\to b\bar{b}$~\cite{BPS}.

In this note we study universality-breaking
effects induced by heavy Majorana neutrinos in leptonic $Z$ decays.
Actually, we will analytically calculate the following physical observable:
\beq
U_{br} \ =\ \frac{\Gamma (Z\to \tau^+\tau^-)\ -\ \Gamma
(Z\to l^+l^-, l=e\ \mbox{or}\ \mu)}{\Gamma (Z\to \tau^+\tau^-)\ +\ \Gamma
(Z\to l^+l^-, l=e\ \mbox{or}\ \mu)}\ .
\eeq 
$U_{br}$ is a measure of universality violation in the leptonic sector
provided lepton masses can be neglected and is experimentally constrained to
be~\cite{PDG}
$ |U_{br}| \leq 1.5\  10^{-2}$.
The model we are concerned with extends the $SM$ by
one right-handed neutrino field for each family. The renormalizable form
of all relevant interactions and details of the notation we will use here can
be found in~\cite{ZPC,KPS}.

In the $SM$ the universality-violating  parameter $U_{br}$
has a value different from zero due to $\tau$ lepton mass effects.
However, pure phase-space~($PS$) corrections turn out to be rather small, i.e.
$|U^{(PS)}_{br}|\ \ \simeq \ \ 3\ [1+(1-4\sin^2\theta_W)^2]^{-1}\
m^2_\tau/M^2_Z  \simeq  1.1\ 10^{-3} $.
This standard source for a non-zero value of $U_{br}$ is beyond the
sensitivity of the present $LEP$ experiments. Since one can expect
to analyze about $10^5$ leptonic decays of the $Z$ boson at $LEP$ per
year,  one may reach an accuracy for $|U_{br}|$ at the level of
$3.\ 10^{-3}$. Since we are interested in values of $|U_{br}|$ much larger
than $|U^{(PS)}_{br}|$, we will neglect lepton-mass effects in the
calculation of quantum corrections.

Decomposing now the transition matrix element ${\cal T}(Z\to l\bar{l})$ into
two parts (with the superscripts (0), (1) denoting the electroweak loop order),
i.e.~${\cal T}(Z\to l\bar{l}) ={\cal T}^{(0)}(Z\to l\bar{l})+
{\cal T}^{(1)}(Z\to l\bar{l})\ $, with
\bmath
{\cal T}^{(0)}(Z\to l\bar{l})\ =\ \frac{ig_W}{4\cos\theta_W}\
\varepsilon_Z^\mu\ \bar{u}_l \gamma_\mu(1-4\sin^2\theta_W -\gamma_5 )
v_{\bar{l}}\ ,
\emath
and defining $\Delta {\cal T}^{(1)} = {\cal T}^{(1)}(Z\to \tau^+ \tau^-)
\ -\ {\cal T}^{(1)}(Z\to l^+ l^-, l \neq \tau)\  $, we find that
\beq
\Delta {\cal T}^{(1)}
\ =\ \frac{ig_W\alpha_W}{8\pi \cos\theta_W} \varepsilon_Z^\mu\ \bar{u}_l
\gamma_\mu (1-\gamma_5 ) v_{\bar{l}}\ \Delta B_{ij} F_Z(\lambda_i,\lambda_j)
\ .
\eeq 
Then, the universality-breaking parameter $U_{br}$ takes the simple form
\beq
U_{br}\ =\ \frac{\mbox{Re}({\cal T}^{(0)}\Delta {\cal T}^{(1)})}
{|{\cal T}^{(0)}|^2}\ =\ \frac{\alpha_W}{\pi}\ \frac{1-2\sin^2\theta_W}
{1+(1-4\sin^2\theta_W)^2} \Delta B_{ij} F_Z(\lambda_i,\lambda_j)\ ,
\eeq 
where
\beq
\lambda_i\ =\ \frac{m^2_{n_i}}{M^2_W}\ ,\qquad
\Delta B_{ij}\ =\ B_{\tau i} B_{\tau j}^\ast\ -
B_{l i} B_{l j}^\ast\ , \ \ \ l\neq \tau\ .
\eeq 
In Eq.~(4), $B_{li}$ is a Cabbibo-Kobayashi-Maskawa-type
$n_G\times 2n_G$ matrix appearing in the leptonic charged-current
interaction,
and $m_{n_i}$ indicates the masses of all neutral leptons $n_i$ in our minimal
scenario (i.e.~$i=1,2,\dots ,2n_G$, with $n_G$ denoting the number of
families). The function $F_Z(\lambda_i,\lambda_j)$
originates from the one-loop graphs depicted in Fig.~1 and contains all the
non-decoupling physics mediated by heavy Majorana neutrinos. Its explicit
form will be discussed below.

Before proceeding to give the analytical form of $F_Z$, we remark
that the universality-violating parameter $U_{br}$ does not involve
any infra-red ($IR$) singularities, as they do not depend on neutrino masses.
Therefore, soft-photon emmision graphs need not to be considered here.
Also, the $Z-l-l$ vertex should be renormalized at the one loop level.
In fact, a large number of renormalization constants should not appear in the
universality-breaking parameter $U_{br}$.
If we adopt the on-shell renormalization scheme~\cite{Sirlin,Hollik}, where the
input renormalization parameters are the electric charge $e$, $M_W$, $M_Z$,
the Higgs mass $M_H$ and all fermion masses contained in the model,
the counterterm
Lagrangian ${\cal L}^C_{int}$ relevant for the renormalization of the
$Z-l-\bar{l}$ vertex is then given by~\cite{Aoki}
\bea
{\cal L}^C_{int}\ &=&\ ie\ Z^{1/2}_{AZ}\, \bar{l}\gamma_\mu l\ Z^{\mu} \ +\
\frac{ie}{4s_W c_W}\ \Bigg[ \ 1\ +\ \frac{\delta e}{e}\
+\ \frac{1-2c^2_W}{2s^2_W}\ \delta\rho\ +\ \delta
Z^{1/2}_{ZZ}\ \nonumber\\
&&+\ \delta Z^l_L \Bigg]\ \bar{l}\gamma_\mu(1-\gamma_5)l\
Z^{\mu}\ -\ \frac{ie\ s_W}{c_W}\ \bar{l} \gamma_\mu \Bigg[
\ 1\ +\ \frac{\delta e}{e}\ +\ \frac{1}{2s^2_W}\ \delta\rho
\ +\ \delta Z^{1/2}_{ZZ}\ \nonumber\\
&&+\ \delta Z^l_L\ \frac{1-\gamma_5}{2}\ +\ \delta Z^l_R\
\frac{1+\gamma_5}{2}\ \Bigg] l\ Z^{\mu},
\eea 
where $\delta \rho =\delta M^2_Z/M^2_Z - \delta M^2_W/M^2_W$, $c_W=
\cos\theta_W = M_W/M_Z$ and $s_W = \sin\theta_W$.
Due to $GIM$-type cancellation~\cite{GIM}
the only non-vanishing contribution to the function $F_Z$ comes from
the wave-function renormalization constants of the left- and right-handed
leptons, i.e.~$\delta Z^l_L$ and $\delta Z^l_R$. In fact, one has
to calculate the difference of the self-energy derivatives given by
$\Delta Z^l_L\ =\ \delta Z^\tau_L\ -\ \delta Z^{l\neq \tau}_L\ =\
\Delta\ \partial\Sigma(\slash{p})/\partial \slash{p} |_{\slash{p}=
m_l,m_\tau\to 0} $.
The corresponding constant for the right-handed leptons $\Delta Z^l_R$ vanishes
in the limit $m_l, m_\tau \to 0$.
It is easy to see that only the neutrino-mass dependent self-energy graphs
mediated by $W^\pm$ and $\chi^\pm$ are of interest here.
The individual contributions to $F_Z(\lambda_i, \lambda_j)$
arising from the diagrams~1(a)--1(f) and those from the wave-function
renormalization constant $\Delta Z^l_L$ are given by
\bea
F_Z^{(a)} & = & \ \frac{1}{2} \bigg[ C_{ij}
\left(\  L_2(\lambda_i,\lambda_j)\ -\
\lambda_Z \Big[ K_1(\lambda_i,\lambda_j)\ -\ K_2(\lambda_i,\lambda_j)\ +\
\tilde{K}(\lambda_i,\lambda_j) \Big] \ \right) \nonumber\\
&&+\ C^\ast_{ij} \sqrt{\lambda_i\lambda_j}
K_1(\lambda_i,\lambda_j)\ \bigg]\ , \\[0.4cm]
F_Z^{(b)} & = & -\ \frac{1}{4} \left[
C_{ij}\lambda_i\lambda_j K_1(\lambda_i,\lambda_j)\  +\  C^\ast_{ij}
\sqrt{\lambda_i\lambda_j}\left( \  \frac{1}{2}\mbox{C}_{UV}\ -\ \frac{1}{2}
\ +\ \lambda_Z\tilde{K}(\lambda_i,\lambda_j)\ \right. \right. \nonumber\\
&&-\ L_2(\lambda_i,\lambda_j)\ \bigg) \bigg]\ , \\[0.4cm]
F_Z^{(c)} & = & -\ \delta_{ij} \left[ \lambda_Z \tilde{I}(\lambda_i)\ +\
3c^2_W L_1(\lambda_i)\right]\ , \\[0.4cm]
F_Z^{(d)} &=& \frac{1}{8}\delta_{ij}(1-2s^2_W)\lambda_i
\left( \ \mbox{C}_{UV}\ -\ 2 L_1(\lambda_i)\ \right)\ ,\\[0.4cm]
F_Z^{(e)}+F_Z^{(f)} & = & -\ \delta_{ij}\ \frac{s^2_W}{c_W}\
\lambda_i I(\lambda_i)\ ,\\[0.4cm]
F_Z^{(\Delta Z_L^l)} & = &-\ \frac{1}{8} \delta_{ij} (1-2s^2_W)
\lambda_i \left( \
\mbox{C}_{UV}\ +\ \frac{3}{2}\ -\ \frac{3}{1-\lambda_i}\ -\
\frac{(\lambda_i+2)\lambda_i\ln\lambda_i}{(1-\lambda_i)^2}\ \right)\ ,
\eea 
where
\bea
C_{ij}\ &=&\ \sum\limits_{k=1}^{n_G} B^\ast_{l_ki} B_{l_kj}\ ,\qquad
\lambda_Z\ =\ \frac{M^2_Z}{M^2_W}\ ,\nonumber\\
{\mbox{C}}_{UV} & = & \frac{1}{\varepsilon} - {\gamma}_E + \ln 4\pi -
\ln \frac{M^2_W}{\mu^2}\ .
\eea 
The functions $I$, $\tilde{I}$, $L_1$, $K_1$, $K_2$, $\tilde{K}$ and
$L_2$ involved in Eqs.~(6)-(11) are given in Appendix~A. It is
straightforward to see that the ultraviolet~($UV$) divergences
(i.e.~${\mbox{C}}_{UV}$) cancel
in the summation of all $F_Z$ terms. To be precise, the $UV$ pole in
$F_Z^{(d)}$ cancels against the $UV$ one of the wave-function
renormalization $F_Z^{(\Delta Z_L^l)}$ and
the $UV$ constant in Eq.~(7) vanishes due to the identity~\cite{AP}:
$\sum\limits_{i=1}^{2n_G} m_{n_i} B_{li}C^{\ast}_{ij}\  =  \ 0 \ $.

For definiteness, we will consider a interfamily-mixing
model with two families only. Employing relations between the mixing
matrix $B_{li}$ and heavy Majorana neutrino masses~\cite{KPS} together with
Eqs.~(3) and (6)--(11), we arrive at the simple result
\beq
|U_{br}|\ \simeq \ \frac{\alpha_W}{8\pi}\ \frac{(s^{\nu_\tau}_L)^4-
(s^{\nu_l}_L)^4}{(1+x^{-1/2})^2}\ \lambda_{N_1}\ \left[
\ 1\ +\ \frac{1}{2}\ln x\ -\ \frac{\ln x}{1-x}\ (1\ +\ 2x^{1/2}) \right]\ ,
\eeq 
where $x=m^2_{N_2}/m^2_{N_1}$.  We also assume $N_2$ to be heavier than
$N_1$, i.e.~$x\geq 1$.  In Eq.~(13) $s^{\nu_l}_L$ is the usual
neutrino-mixing angle between heavy Majorana neutrinos and the charged lepton
$l$ and is generally defined as~\cite{Lang} $(s^{\nu_l}_L)^2 =
\sum\limits_{N_i} |B_{lN_i}|^2\ $.
The neutrino-mixing angle $(s^{\nu_\tau}_L)^2$ turns out to be severely
constrained by the recent $LEP$
data on $\tau$ decays. In fact, a global analysis allowing mixing of exotic
particles gives an upper bound of about $7.\ 10^{-2}$~\cite{Nardi}.
The $e$- and $\mu$-family is much more constrained,
i.e.~$(s^{\nu_e(\mu)}_L)^2 < 0.01$. Due to this fact we have to deal with
universality-breaking effects in the heaviest lepton family only. It is
important to notice that the non-decoupling terms (i.e.~proportional to
$m^2_N/M^2_W$) come from the "seemingly" suppressed $(s^{\nu_\tau}_L)^4$
terms. Table~1 shows the dramatic non-decoupling behaviour of
the loop function $F_Z$. For comparison, we also show the corresponding
values for a calculation where terms proportional to $(s^{\nu_\tau}_L)^4$
have been neglected. In our numerical estimates we have assumed that there
is no large mass difference between the two heavy Majorana neutrinos,
i.e.~$x\simeq 1$. We have varied the heavy neutrino mass
$m_N$ $(\sim m_{N_1} \sim m_{N_2})$ up to its perturbative unitarity bound.
Such an upper bound may be imposed by requiring that the total width of $N_1$
and $N_2$, denoted by $\Gamma_{N_i}$, satisfies the condition
$\Gamma_{N_i}/m_{N_i} \leq 1/2$~\cite{KPS}. This leads to the constraint on
the mass of the lightest heavy Majorana neutrino $N_1$
\beq
m^2_{N_1} \ \leq \ \frac{2M^2_W}{\alpha_W(s^{\nu_\tau}_L)^2}\
\frac{1\ +\ x^{-1/2}}{x^{1/2}}\ .
\eeq 
Taking the above upper bound into account, we find that the
universality-violating parameter $|U_{br}|$ can be up to 10 times larger
than the naive value obtained by considering only terms proportional to
the mixing $(s^{\nu_\tau}_L)^2$. This enhancement factor
(i.e.~the crucial $m^2_N/M^2_W$ dependence)
results from the coupling of the would-be charged Goldstone boson $\chi^+$ to
the heavy Majorana neutrinos in the diagram~1(b). If we assume very large mass
differences between the heavy neutrinos, $|U_{br}|$ smoothly decreases to
negligibly small values. The reason is that one effectively recovers the one
generation "see-saw" model in such a case, i.e.~$m_{N_1}\to 0$ as $x \to
\infty$ in Eq.~(14). In Fig.~(2) we present exclusion plots for
$LEP$ experiments.
We see, for example, that possible universality-breaking effects of the
order of~$10^{-2}$ can easily be understood within our minimal model.

Non-$SM$ contributions, due to heavy neutrinos, could also be observed by
comparing the prediction of the $SM$ for the leptonic rates $\Gamma( Z
\to l^+ l^-)$ with those epxected when one includes heavy neutral lepton
effects. Besides the universality-violating phenomena discussed in this
paper, one could have non-decoupling physics introduced by these heavy
neutral leptons which can be constrained by analyzing the oblique
electroweak parameters $S$, $T$, $U$ or $\varepsilon_1$, $\varepsilon_2$,
$\varepsilon_3$ as defined in~\cite{Peskin}. These contributions occur both
through vacuum polarization terms and in vertex corrections universal for
all charged lepton flavors. However, a first estimate for the $\delta\rho$
parameter~\cite{Velt} (i.e.~$\delta\rho=\alpha_{em} T$)
provides a weaker bound than that obtained by Eq.~(14). In particular,
the dominant non-$SM$ contribution (denoted below as $\delta\rho^\ast$)
to $\delta\rho$, apart from heavy-top and Higgs-particle effects, comes from
the $\nu_iN_j$, $N_iN_j$ intermediate states of the $Z$ self-energy graph.
In this way, one gets
\beq
\delta\rho^\ast|_{q^2=0}\ \simeq \ \frac{\alpha_W}{16\pi}
\ (s^{\nu_\tau}_L)^4\ \frac{m^2_N}{M^2_W}\ ,\  \  \mbox{for}\ \ x=1\ .
\eeq 
We can readily compare Eq.~(15) with constraints on the masses of the heavy
neutrinos that are derived on the basis of perturbative unitarity. For example,
if $(s^{\nu_\tau}_L)^2=0.1$, the maximal value that $\delta\rho^\ast$ can
take is 0.8~$10^{-2}$, which is still in accordance with
phenomenological contraints of a possible mass shift of the
$W$ boson~\cite{Kniehl}. However, to complete the analysis a global
consideration of all electroweak oblique corrections is required.

In conclusion, we have explicitly shown that heavy neutral leptons
introduce a quadratic mass dependence (i.e.~$\alpha_W\, m^2_N/M^2_W$) in the
leptonic vertex function $Z-l-\bar{l}$. This situation is not peculiar for
the minimal model considered in this work, but a general feature for all
theories based on the spontaneous symmetry-breaking mechanism,  e.g.
similar effects will be present in the
model described in~\cite{JB}. In general, we have found that
the mass of possible non-decoupling neutral
particles can be constrained by the already existing or future
$LEP$ data, as a function of their mixing to the ordinary charged leptons.

$AP$ wishes to thank University of Val\`encia for
the kind hospitality. Helpful discussions and criticisms of F.J.~Botella,
B.~Kniehl, A.~Santamaria and J.W.F.~Valle are gratefully acknowledged.
The work of $JB$ has been supported by CICYT, {\em Spain}, under Grant
No.~AEN 90-0040,
the work of $JGK$ and $KS$ by the BMFT, $FRG$, under the contract No.~06MZ730,
and the work of $AP$ by a grant from the Postdoctoral Graduate College of
Mainz, $FRG$.

\setcounter{section}{1}
\setcounter{equation}{0}
\def\theequation{\Alph{section}\arabic{equation}}

\begin{appendix}
\section{The one-loop integrals}
\indent

It is useful first to define the functions
$B_1(\lambda_i)$ and $B_2(\lambda_i,\lambda_j)$ which are given by
the following expressions:
\bea
B_1(\lambda_i) &\  = \ & (1-y)\lambda_i + y[1-\lambda_Z\; yx(1-x)]\ , \\
B_2(\lambda_i,\lambda_j) &\ = \ & 1-y+y[x\lambda_i+(1-x)\lambda_j-
\lambda_Z\; yx(1-x)] \ ,
\eea 
where $x$ and $y$ are Feynman parameters.
The loop functions $I$, $\tilde{I}$, $L_1$, $K_1$, $K_2$, $\tilde{K}$ and
$L_2$ can then be written in terms of the following integrals:
\bea
I(\lambda_i)\ &=&\ \int \frac{dxdy\, y}{B_1(\lambda_i)}\ ,\\
\tilde{I}(\lambda_i)\ &=&\ \int \frac{dxdy\, y^2}{B_1(\lambda_i)}[1-yx(1-x)]
\ ,\\
L_1(\lambda_i)\ &=&\ \int dxdy\ y\ln B_1(\lambda_i)\ ,\\
K_1(\lambda_i,\lambda_j)\ &=&\ \int\frac{dxdy\ y}{B_2(\lambda_i,\lambda_j)}
\ ,\\
K_2(\lambda_i,\lambda_j)\ &=&\ \int\frac{dxdy\ y^2}{B_2(\lambda_i,\lambda_j)}
\ ,\\
\tilde{K}(\lambda_i,\lambda_j)\  ,\
&=&\ \int\frac{dxdy\ y^3x(1-x)}{B_2(\lambda_i,\lambda_j)}\ ,\\
L_2(\lambda_i,\lambda_j)\ &=&\ \int dx dy \ y\ln B_2(\lambda_i,\lambda_j)\ ,
\eea 
where the integration variables $x$ and $y$ are constrained to the interval
$[0,1]$. The above integrals can be best performed numerically.

\end{appendix}

\newpage

\centerline{\bf\Large Figure and Table Captions }
\vspace{1cm}
\newcounter{fig}
\begin{list}{\bf\rm Fig. \arabic{fig}: }{\usecounter{fig}
\labelwidth1.6cm \leftmargin2.5cm \labelsep0.4cm \itemsep0ex plus0.2ex }

\item One-loop irreducible vertex graphs contributing to the
non-universality parameter $U_{br}$ in the Feynman--'t Hooft gauge.

\item {\em Exclusion plots for $LEP$ experiments}.
The areas lying to the right of the curves are excluded due to the
following conditions: (i) The validity of perturbative unitarity (solid
line) -- see also text, (ii) $|U_{br}| \leq 1.\ 10^{-2}$ (dashed line),
(iii) $|U_{br}| \leq 7.\ 10^{-3}$ (dash-dotted line),
(iv) $|U_{br}| \leq 3.\ 10^{-3}$ (dotted line).

\end{list}
\newcounter{tab}
\begin{list}{\bf\rm Tab. \arabic{tab}: }{\usecounter{tab}
\labelwidth1.6cm \leftmargin2.5cm \labelsep0.4cm \itemsep0ex plus0.2ex }

\item  Numerical estimates for the universality-violating quantity $|U_{br}|$
within the perturbatively allowed parameter space. The values in
parentheses are obtained by neglecting the "seemingly" suppressed terms
proportional to $(s^{\nu_\tau}_L)^4$.

\end{list}

\newpage

\centerline{\bf\Large Table 1}
\vspace{0.6cm}
\hoffset=+1.cm
\begin{tabular*}{16.95cm}{|c|ccccccc|}
\hline
 &&&&$(s^{\nu_\tau}_L)^2$&&& \\
$m_N$ & 0.06 & 0.05 & 0.04 & 0.03 & 0.02 & 0.01 & 0.005 \\
$[$TeV$]$ &&&&&&& \\
\hline\hline
0.3&1.7 $10^{-4}$&1.3 $10^{-4}$&9.8 $10^{-5}$&6.7 $10^{-5}$&4.0 $10^{-5}$&
                                              1.8 $10^{-5}$&8.4 $10^{-6}$ \\
&(9.4 $10^{-5}$)&(7.8 $10^{-5}$)&(6.3 $10^{-5}$)&(4.7 $10^{-5}$)&
                    (3.1 $10^{-5}$)&(1.6 $10^{-5}$)&(7.8 $10^{-6}$) \\
\hline
0.5&4.9 $10^{-4}$&3.8 $10^{-4}$&2.8 $10^{-4}$&1.9 $10^{-4}$&1.2 $10^{-4}$&
                                              5.2 $10^{-5}$&2.5 $10^{-5}$ \\
&(2.8 $10^{-4}$)&(2.3 $10^{-4}$)&(1.8 $10^{-4}$)&(1.4 $10^{-4}$)&
                 (9.2 $10^{-5}$)&(4.6 $10^{-5}$)&(2.3 $10^{-5}$) \\
\hline
0.7&8.4 $10^{-4}$&6.4 $10^{-4}$&4.7 $10^{-4}$&3.2 $10^{-4}$&1.9 $10^{-4}$&
                                              8.2 $10^{-5}$&3.8 $10^{-5}$ \\
&(4.2 $10^{-4}$)&(3.5 $10^{-4}$)&(2.8 $10^{-4}$)&(2.1 $10^{-4}$)&
                 (1.4 $10^{-4}$)&(7.0 $10^{-5}$)&(3.5 $10^{-5}$) \\
\hline
1.0&1.4 $10^{-3}$&1.1 $10^{-3}$&7.7 $10^{-4}$&5.1 $10^{-4}$&2.9 $10^{-4}$&
                                              1.2 $10^{-4}$&5.5 $10^{-5}$ \\
&(5.9 $10^{-4}$)&(4.9 $10^{-4}$)&(3.9 $10^{-4}$)&(2.9 $10^{-4}$)&
                 (1.9 $10^{-4}$)&(9.8 $10^{-5}$)&(4.9 $10^{-5}$)  \\
\hline
2.0&4.3 $10^{-3}$&3.1 $10^{-3}$&2.1 $10^{-3}$&1.3 $10^{-3}$&6.9 $10^{-4}$&
                                              2.5 $10^{-4}$&1.0 $10^{-4}$ \\
&(9.4 $10^{-4}$)&(7.8 $10^{-4}$)&(6.3 $10^{-4}$)&(4.7 $10^{-4}$)&
                 (3.1 $10^{-4}$)&(1.5 $10^{-4}$)&(7.8 $10^{-5}$) \\
\hline
3.0&8.7 $10^{-3}$&6.2 $10^{-3}$&4.1 $10^{-3}$&2.5 $10^{-3}$&1.2 $10^{-3}$&
                                              4.0 $10^{-4}$&1.5 $10^{-4}$ \\
&(1.1 $10^{-3}$)&(9.6 $10^{-4}$)&(7.7 $10^{-4}$)&(5.7 $10^{-4}$)&
                 (3.8 $10^{-4}$)&(1.9 $10^{-4}$)&(9.6 $10^{-5}$) \\
\hline
4.0 &1.5 $10^{-2}$&1.0 $10^{-2}$&6.8 $10^{-3}$&4.0 $10^{-3}$&1.9 $10^{-3}$&
                                               5.9 $10^{-4}$&2.0 $10^{-4}$ \\
&(1.3 $10^{-3}$)&(1.1 $10^{-3}$)&(8.7 $10^{-4}$)&(6.5 $10^{-4}$)&
                 (4.3 $10^{-4}$)&(2.2 $10^{-4}$)&(1.1 $10^{-4}$) \\
\hline
6.0& $-$ & $-$ & $-$ &8.3 $10^{-3}$&3.9 $10^{-3}$&1.1 $10^{-3}$&3.3 $10^{-4}$\\
&&&&(7.5 $10^{-4}$)&(5.0 $10^{-4}$)&(2.5 $10^{-4}$)&(1.2 $10^{-4}$) \\
\hline
8.0& $-$ & $-$ & $-$ & $-$ & $-$ & 1.8 $10^{-3}$&5.1 $10^{-4}$ \\
&&&&&&(2.7 $10^{-4}$)&(1.4 $10^{-4}$) \\
\hline
10.0& $-$ & $-$ & $-$ & $-$ & $-$ & $-$ &7.3 $10^{-4}$ \\
&&&&&&&(1.4 $10^{-4}$) \\
\hline
\end{tabular*}

\hoffset=-1.cm

\end{document}